\documentclass[a4paper,12pt]{article}

\usepackage[english]{babel}
\usepackage{authblk}
\usepackage{dsfont}
\usepackage{amsfonts}
\usepackage{mathrsfs}
\usepackage{amssymb}        
\usepackage{amsmath}
\usepackage{graphicx,caption}
\usepackage{float}
\usepackage[a4paper]{geometry} 
\usepackage{verbatim}
\usepackage{pstricks}
\usepackage{amsthm}
\usepackage{mathrsfs}
\usepackage{enumitem} 
\usepackage{xcolor}
\usepackage{bm}

\usepackage{graphicx}
\usepackage{subcaption}

\usepackage{hyperref}
\usepackage{cleveref}
\hypersetup{
    colorlinks=true,
    urlcolor=blue,
    citecolor=red
}

\usepackage{tikz}
\usetikzlibrary{arrows.meta,decorations.pathmorphing}

\begin{document}

\title{Many-body localization:\\ an introduction for mathematicians}
\author[*]{Francois Huveneers}
\affil[*]{Department of Mathematics, King’s College London, Strand, London WC2R 2LS, United Kingdom}
\date{\today}

\maketitle

\begin{abstract}
These notes are based on a mini-course delivered at the Les Houches
Summer School, \emph{Quantum Theory on All Scales}, in August 2026.
They introduce many-body localization to mathematicians, emphasizing
the physical mechanisms that support or destabilize it.
The mathematical discussion deals with how perturbative rotations, renormalization, and probabilistic estimates are used in a rigorous approach to many-body localization.
\end{abstract}

\section{Introduction}

What is many-body localization (MBL)? It is a striking example of
ergodicity breaking, and it challenges the usual picture of quantum
statistical mechanics. We normally expect an isolated many-body system
to approach thermal equilibrium: at long times, local observables
should depend only on a few conserved quantities, such as the total
energy, rather than on the details of the initial state. An MBL system
retains local memory of its initial state even at arbitrarily long
times.

Historically, the question arose from transport. Anderson showed that
disorder can prevent transport in a noninteracting
system~\cite{Anderson1958}. Can this localization survive interactions,
which allow particles to exchange energy? Foundational works by
Gornyi, Mirlin and Polyakov~\cite{Gornyi2005} and Basko, Aleiner and
Altshuler~\cite{Basko2006} addressed this question. The focus then
broadened from the absence of transport to the failure of
thermalization, notably through the numerical study of Oganesyan and
Huse~\cite{Oganesyan2007}. We will adopt this broader viewpoint:
MBL is not just about whether energy or particles can move, but about
whether the system can reach thermal equilibrium.

We will first describe MBL at a physical and a theoretical level
(Section~\ref{sec: what is MBL}). In particular, we will explain how
quasi-local conserved quantities can preserve memory of the initial
state. We will then ask what makes localization possible and what can
destabilize it, focusing on resonances and thermal avalanches
(Section~\ref{sec: pro and con MBL}). Finally, we will discuss how to
tackle MBL mathematically, starting with the restricted problem of
localization in a resonance-free region
(Section~\ref{sec: math MBL}).
These notes do not aim to survey the literature. For a broader
discussion and citations, we refer to the review~\cite{Sierant2025}.

An important preliminary remark: MBL concerns closed systems.
The failure to thermalize occurs under the system's own dynamics,
without coupling to an environment. This isolation is essential.
We may later introduce a coupling to a bath, but only as a way to
probe the system's own dynamics.

\section{What is MBL?}\label{sec: what is MBL}

We describe MBL systems by contrasting them with thermalizing,
or ergodic, systems. We then develop a theoretical picture in
terms of local integrals of motion and conclude with commonly
used numerical probes of MBL.

\subsection{Ergodicity breaking: Intuitive physical picture}

Consider two rods of the same material, initially separated and
prepared at temperatures $T_h>T_c$. Each rod is initially in thermal
equilibrium. We then bring them into contact, keeping the combined
system isolated from the environment.

\begin{center}
\begin{tikzpicture}[x=1cm,y=1cm]
    \filldraw[fill=blue!0,draw=black,rounded corners=2pt]
        (0,0) rectangle (4,0.5);
    \filldraw[fill=blue!0,draw=black,rounded corners=2pt]
        (4.2,0) rectangle (8.2,0.5);
    \node at (2,0.25) {$T_h$};
    \node at (6.2,0.25) {$T_c$};
\end{tikzpicture}
\end{center}

What happens after a long time? For an ergodic, or thermalizing,
material, energy flows between the rods until their temperatures
become equal. The final temperature is fixed by the total energy.
The system has lost the memory of where the energy was initially
stored.

An MBL material behaves differently. Energy can be exchanged near
the contact, but this does not lead to equilibration of the two rods.
Far from the contact, the hot and cold regions retain their initial
temperatures (the temperature may no longer be well defined near the contact point). 
The system thus keeps a local memory of its preparation,
even at long times.

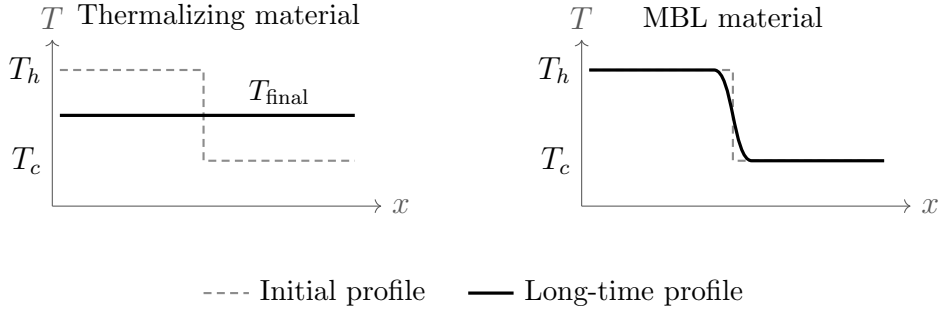
\begin{center}
\begin{tikzpicture}[
    x=1cm,y=1cm,
    initial/.style={black!45,densely dashed,thick},
    final/.style={black,very thick},
    axis/.style={->,black!65,thin}
]
    \begin{scope}
        \node[font=\small] at (2.2,2.5) {Thermalizing material};

        \draw[axis] (0,0) -- (4.35,0) node[right] {$x$};
        \draw[axis] (0,0) -- (0,2.2) node[above] {$T$};

        \node[left] at (0,1.8) {$T_h$};
        \node[left] at (0,0.6) {$T_c$};

        \draw[initial]
            (0.1,1.8) -- (2,1.8) -- (2,0.6) -- (4,0.6);
        \draw[final] (0.1,1.2) -- (4,1.2);
        \node[above,font=\small] at (3,1.2) {$T_{\mathrm{final}}$};

    \end{scope}

    \begin{scope}[xshift=7cm]
        \node[font=\small] at (2,2.5) {MBL material};

        \draw[axis] (0,0) -- (4.35,0) node[right] {$x$};
        \draw[axis] (0,0) -- (0,2.2) node[above] {$T$};

        \node[left] at (0,1.8) {$T_h$};
        \node[left] at (0,0.6) {$T_c$};


        \draw[initial]
            (0.1,1.8) -- (2,1.8) -- (2,0.6) -- (4,0.6);
        \draw[final]
            (0.1,1.8) -- (1.75,1.8)
            .. controls (2,1.8) and (2,0.6) .. (2.25,0.6)
            -- (4,0.6);

    \end{scope}

    \draw[initial] (2,-1.15) -- (2.6,-1.15);
    \node[right,font=\small] at (2.6,-1.15) {Initial profile};
    \draw[final] (5.5,-1.15) -- (6.1,-1.15);
    \node[right,font=\small] at (6.1,-1.15) {Long-time profile};
\end{tikzpicture}
\end{center}


We can phrase this in more fundamental terms. 
A thermalizing system locally approaches the maximum-entropy state
compatible with its energy and other conserved quantities. This is
the thermodynamic prediction. In an MBL system, additional constraints
prevent it from reaching that state. What are these constraints,
and why do they survive interactions? These are the questions to be addressed.

\subsection{Ergodicity breaking: Theoretical picture}

We will work with spin-$1/2$ chains throughout. For $L$ spins, the
Hilbert space is
\[
    \mathcal H=(\mathbb C^2)^{\otimes L},
    \qquad d=\dim\mathcal H=2^L.
\]
The Hamiltonian is a sum of local terms,
\[
    H=\sum_{i=1}^L H_i.
\]
Here locality means that $H_i$ acts nontrivially only on a fixed
number of spins near site $i$, independent of $L$. We also assume
that the norms $\|H_i\|$ are uniformly bounded. We write
\[
    Z_i=\mathrm{Id}\otimes\cdots\otimes Z\otimes\cdots
    \otimes\mathrm{Id},
\]
where the Pauli matrix $Z$ acts on the $i$-th factor. The Pauli operators
$X_i$ and $Y_i$ are defined in the same way.

\paragraph{Quantum typicality.}
Quantum typicality is a
mathematical statement about generic states, independent of any
Hamiltonian or dynamics~\cite{Goldstein2006,Popescu2006}.
If $A$ is a fixed local observable and $|\psi\rangle$ is a normalized
state chosen uniformly at random, according to the Haar measure, then
with high probability,
\[
    \langle\psi|A|\psi\rangle\simeq
    \langle A\rangle_\infty=\frac{\operatorname{Tr}(A)}{d}.
\]
The right-hand side is the expectation in the infinite-temperature
state $\rho_\infty=\mathrm{Id}/d$. More precisely, the Haar mean
and variance satisfy~\cite{Reimann2008}
\[
    \mathbb E_\psi\bigl[\langle\psi|A|\psi\rangle\bigr]
    =\frac{\operatorname{Tr}(A)}{d},
\]
and
\[
    \mathbb E_\psi\left[
      \left|\langle\psi|A|\psi\rangle-\langle A\rangle_\infty\right|^2
    \right]
    =
    \frac{\operatorname{Tr}(A^2)/d
      -\bigl(\operatorname{Tr}(A)/d\bigr)^2}{d+1}
    \leq\frac{\|A\|^2}{d+1}.
\]
For a local observable with bounded norm, this variance is
exponentially small in $L$.

\paragraph{Eigenstate Thermalization Hypothesis (ETH).}
Eigenstates of a local Hamiltonian are not chosen at random, but typicality still describes them.
This is the content of the Eigenstate Thermalization Hypothesis
(ETH)~\cite{Deutsch1991,Srednicki1994,Rigol2008}. We will mostly
consider infinite temperature, whose mean energy is
\[
    E_\infty=\frac{\operatorname{Tr}(H)}{d}.
\]
For eigenstates with energy density near $E_\infty/L$, ETH predicts
\[
    \langle E|A|E\rangle\simeq
    \langle A\rangle_\infty=\frac{\operatorname{Tr}(A)}{d}
\]
for local observables $A$. In this local sense, the eigenstates
behave like typical states. Unlike quantum typicality, this is
a hypothesis about the eigenvectors of a Hamiltonian. It is
supported by extensive evidence for nonintegrable thermalizing
systems, but there is no general proof for such local Hamiltonians.

For eigenstates at an energy density corresponding to a finite
temperature $T$, the thermal average should instead be
$\langle A\rangle_T$. If additional conserved quantities are
present, the average must also account for their values.

To see why ETH is necessary for thermalization, imagine preparing
the system in an eigenstate $|E\rangle$ (though this may not be easy in practice). 
Its time evolution only adds a phase,
\[
    e^{-iHt}|E\rangle=e^{-iEt}|E\rangle,
\]
so every expectation value remains constant:
\[
    \langle E|A(t)|E\rangle=\langle E|A|E\rangle.
\]
Therefore, the convergence of $\langle E|A(t)|E\rangle$ to the thermal average is only possible if ETH is satisfied.

\subsection{A simple example of an MBL Hamiltonian}

Consider
\[
    H=\sum_{i=1}^L h_iZ_i,
\]
with independent random fields $h_i$ drawn from a uniform distribution in $[-W,W]$.
To avoid clutter, we measure energies in units of the disorder strength, setting $W=1$ throughout unless stated otherwise.
The eigenstates of $H$ are classical product states:
\[
    |\bm s \rangle = |s_1,\ldots,s_L\rangle
    =|s_1\rangle\otimes\cdots\otimes|s_L\rangle,
    \qquad s_i=\pm1,
\]
where $Z|s_i\rangle=s_i|s_i\rangle$. Each spin therefore has a
definite orientation, up or down, along the $z$ axis, and there is
no entanglement between spins. 
We call these states classical because they are
labelled by ordinary spin configurations. Their energies are
\[
    E(\bm s)=\sum_{i=1}^L h_is_i.
\]
For a finite chain, the disorder makes the spectrum nondegenerate
with probability one.

ETH is violated: for example,
\[
    \langle E|Z_i|E\rangle=\pm1,
    \qquad \langle Z_i\rangle_\infty=0.
\]
In particular, eigenstates near the infinite-temperature energy density
do not reproduce the thermal value. 
This Hamiltonian can be taken as the simplest example of an MBL Hamiltonian. 
Note that MBL concerns the bulk of the spectrum, rather than only states near the ground state.

The essential question is robustness. In this model, every $Z_i$
is exactly conserved, so the absence of thermalization is immediate.
But this alone does not establish a stable phase: the model is
fine-tuned and integrable, and generic perturbations destroy these
simple conservation laws. MBL posits that localization and the failure
of thermalization nevertheless survive sufficiently weak, generic
local perturbations~\cite{Gornyi2005,Basko2006,Oganesyan2007,Imbrie2016}.
The conserved quantities may change, but local memory persists.
It is this stability over a range of Hamiltonians that makes MBL
more than an isolated example of nonthermal behavior.

For example, consider
\begin{equation}\label{eq: main hamiltonian}
    H=\sum_{i=1}^L h_iZ_i
      +J\sum_{i=1}^{L-1} Z_iZ_{i+1}
      +\gamma\sum_{i=1}^L X_i
\end{equation}
where $h_i$ are the same random fields as before. 
The interaction term couples neighboring spins, while the transverse
field flips spins and breaks the conservation of the individual $Z_i$.
At fixed nonzero $J$, the expected transition out of the
localized regime is shown schematically below:

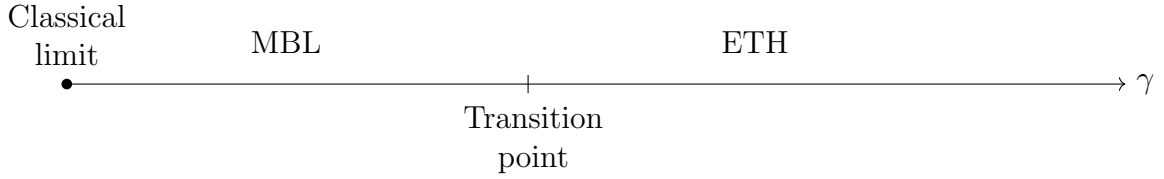
\begin{center}
\begin{tikzpicture}[x=1cm,y=1cm]
    \draw[->] (0,0) -- (14,0) node[right] {$\gamma$};

    \fill (0,0) circle (2pt);
    \node[align=center] at (0,0.65)
        {Classical\\limit};

    \node at (2.9,0.55) {MBL};

    \draw (6.1,-0.12) -- (6.1,0.12);
    \node[align=center,anchor=north west] at (5.1,-0.15)
        {Transition\\point};

    \node at (9.1,0.55) {ETH};
\end{tikzpicture}
\end{center}
As emphasiezd above, MBL is understood as a stability property. 
This stability is not expected to persist arbitrarily far from the classical limit (or non-interacting limit in most physical papers). 
The diagram above sketches a transition to a thermalizing regime as $\gamma$ increases, but
we are taking no stance on what happens at very large $\gamma$.

\subsection{A misleading theoretical picture}

A tempting idea is to view MBL as Anderson localization in Fock space.
Represent the classical configurations of the chain as the vertices of
an $L$-dimensional hypercube, with edges joining configurations that
differ by a single spin flip. For three spins, this gives a cube:

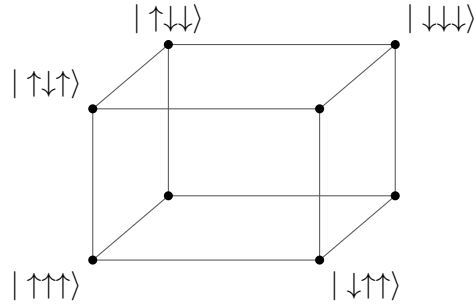
\begin{center}
\begin{tikzpicture}[scale=1, every node/.style={font=\small}]
    \coordinate (A) at (0,0);
    \coordinate (B) at (3,0);
    \coordinate (C) at (3,2);
    \coordinate (D) at (0,2);

    \coordinate (E) at (1,0.85);
    \coordinate (F) at (4,0.85);
    \coordinate (G) at (4,2.85);
    \coordinate (H) at (1,2.85);

    \draw[black!65] (A)--(B)--(C)--(D)--cycle;
    \draw[black!65] (E)--(F)--(G)--(H)--cycle;
    \draw[black!65] (A)--(E) (B)--(F) (C)--(G) (D)--(H);

    \foreach \p in {A,B,C,D,E,F,G,H}
        \fill (\p) circle (1.7pt);

    \node[below left] at (A) {$|\uparrow\uparrow\uparrow\rangle$};
    \node[below right] at (B) {$|\downarrow\uparrow\uparrow\rangle$};
    \node[above left] at (D) {$|\uparrow\downarrow\uparrow\rangle$};
    \node[above] at (H) {$|\uparrow\downarrow\downarrow\rangle$};
    \node[above right] at (G) {$|\downarrow\downarrow\downarrow\rangle$};
\end{tikzpicture}
\end{center}

On this hypercube, the Hamiltonian in \eqref{eq: main hamiltonian} has the form
\[
    H=V+\gamma\Delta,
\]
where $V$ is diagonal and $\Delta$ is the adjacency operator.
This resembles the one-particle Anderson model: a random potential
on the vertices, with hopping between neighboring vertices.
For one-particle Anderson localization at strong disorder, an
eigenstate is typically concentrated near a single site, with small
tails on other sites, apart from occasional resonances. One could have
therefore expected a similar picture here: eigenstates close to
individual classical states, with appreciable weight on only a few
configurations.

But this is not the right many-body picture. Even weak mixing at
each site can accumulate over the whole chain. MBL eigenstates can
therefore spread over exponentially many classical configurations,
of order $e^{cL}$, while remaining localized in a physical
sense~\cite{Bauer2013}.

\subsection{A plausible theoretical picture} 

Since $H$ is Hermitian, there always exists a unitary $U$ such that
\[
    U^\dagger H U=D,
\]
with $D$ diagonal in the classical basis. 
The essential requirement for localization is that $U$ can be chosen to preserve locality: conjugation by $U$ and by $U^\dagger$ must take local observables to quasi-local ones.
This is the theoretical definition of MBL that we will use
~\cite{Serbyn2013,Huse2014,Imbrie2016}.

Schematically, saying that $U$ maps a local observable $O$ to a quasi local one means that there exist constants $C,\xi<+\infty$ such that 
\[
    U^\dagger O U
    =\sum_{\substack{I\text{ interval}:\\
        I\cap\operatorname{supp}(O)\neq\emptyset}}O_I,
    \qquad \operatorname{supp}(O_I)\subseteq I,
\]
with
\[
    \|O_I\|\leq Ce^{-|I|/\xi}\|O\|.
\]
A similar condition holds for $UOU^\dagger$. These locality estimates
must remain meaningful as the system size grows. 
In a realistic disordered system, $C$ cannot truly be taken to be a constant; it must be a random variable that depends on the location of $O$ and on the disorder realization.

This definition describes a fully MBL system: the same quasi-local
change of basis describes all eigenstates of the Hamiltonian,
\[
    |E_{\bm s}\rangle=U|s_1,\ldots,s_L\rangle, \qquad \bm s = (s_1,\ldots,s_L).
\]
All eigenstates are therefore localized. We will not discuss here whether other forms of MBL may exist or not.

We can then construct local integrals of motion (LIOMs),
\[
    \widetilde Z_i=UZ_iU^\dagger,
    \qquad 1\leq i\leq L.
\]
Other choices are possible~\cite{Ros2015}. These operators have the
spectrum and trace of spins,
\[
    \sigma(\widetilde Z_i)=\{-1,+1\},
    \qquad \operatorname{Tr}(\widetilde Z_i)=0.
\]
They are conserved and commute with one another:
\[
    [\widetilde Z_i,H]
    =U[Z_i,U^\dagger HU]U^\dagger=U[Z_i,D]U^\dagger=0,
\]
\[
    [\widetilde Z_i,\widetilde Z_j]
    =U[Z_i,Z_j]U^\dagger=0.
\]
They also form a complete set: every operator diagonal in the eigenbasis
selected by $U$ is a linear combination of products of the
$\widetilde Z_i$, including the identity.

So far, the properties of the $\widetilde Z_i$ are completely generic and could be obtained for any Hamiltonian on this Hilbert space. 
The crucial extra property is that the $\widetilde Z_i$ are quasi-local near their respective sites: 
\[
    \widetilde Z_i = 
    U Z_i U^\dagger
    =\sum_{I\ni i,\text{ interval}}O_I,
    \qquad \operatorname{supp}(O_I)\subseteq I,
    \qquad \|O_I \| \le C e^{-|I|/\xi}.
\]
The $\widetilde Z_i$ are called local integrals of motion (LIOMs). 

Together, these properties give an emergent form of integrability.
They also explain the violation of ETH. In each eigenstate,
\[
    \langle E_{\boldsymbol s}|\widetilde Z_i|E_{\boldsymbol s}\rangle
    =s_i=\pm1,
\]
while 
\[
    \langle\widetilde Z_i\rangle_\infty= \langle Z_i \rangle_\infty =  0.
\]
Although $\widetilde Z_i$ is quasi-local rather than strictly local,
it can be approximated in operator norm by an observable supported
in a sufficiently large, fixed neighborhood of $i$.

\subsection{Practical numerical diagnostics of MBL}\label{sec: numerical probe MBL}

We describe two popular numerical probes of MBL. 

\paragraph{Level statistics.}
A first diagnostic is the statistics of nearby energy levels~\cite{Oganesyan2007}. Order the
many-body spectrum as $E_1\leq E_2\leq\cdots$:

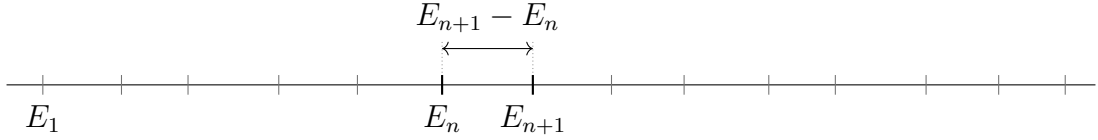
\begin{center}
\begin{tikzpicture}[x=0.8cm,y=1cm]
    \draw (0,0) -- (18,0);

    \foreach \x in {0.6,1.9,3.0,4.5,5.8,
                    10.0,11.2,12.6,13.7,15.2,16.4,17.5}
        \draw[black!55] (\x,-0.13) -- (\x,0.13);

    \draw[thick] (7.2,-0.13) -- (7.2,0.13);
    \draw[thick] (8.7,-0.13) -- (8.7,0.13);

    \node[below] at (0.6,-0.13) {$E_1$};
    \node[below] at (7.2,-0.13) {$E_n$};
    \node[below] at (8.7,-0.13) {$E_{n+1}$};

    \draw[black!40,densely dotted]
        (7.2,0.16) -- (7.2,0.6);
    \draw[black!40,densely dotted]
        (8.7,0.16) -- (8.7,0.6);
    \draw[<->] (7.2,0.48) -- (8.7,0.48);
    \node[above] at (7.95,0.55) {$E_{n+1}-E_n$};
\end{tikzpicture}
\end{center}
Define the normalized gaps by
\[
    s_n=\frac{E_{n+1}-E_n}{\langle\Delta E\rangle},
\]
where $\langle\Delta E\rangle$ is the mean gap. In the MBL regime,
we expect Poisson statistics, with no level repulsion, as distant parts
of the system have little influence on one another. In the thermal
regime, we expect random-matrix statistics with level repulsion. For
the real Hamiltonians considered here, the relevant ensemble is the
Gaussian orthogonal ensemble (GOE).

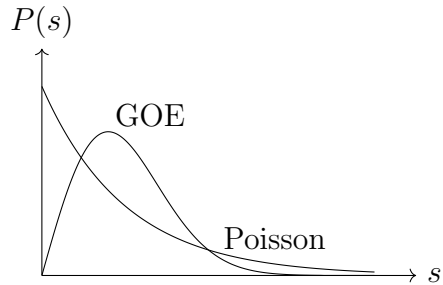
\begin{center}
\begin{tikzpicture}[x=1.1cm,y=2.5cm]
    \draw[->] (0,0) -- (4.5,0) node[right] {$s$};
    \draw[->] (0,0) -- (0,1.2) node[above] {$P(s)$};

    \draw[domain=0:4,samples=100,smooth]
        plot (\x,{pi/2*\x*exp(-pi*\x*\x/4)});
    \node at (1.3,0.85) {GOE};

    \draw[domain=0:4,samples=100,smooth]
        plot (\x,{exp(-\x)});
    \node at (2.8,0.20) {Poisson};
\end{tikzpicture}
\end{center}

A convenient number is the ratio of consecutive gaps,
\[
    r_n=\frac{\min(E_{n+1}-E_n,E_n-E_{n-1})}
               {\max(E_{n+1}-E_n,E_n-E_{n-1})}.
\]
Its mean is approximately
\[
    \langle r\rangle_{\mathrm{GOE}}\simeq0.53,
    \qquad \langle r\rangle_{\mathrm{Poisson}}\simeq0.39.
\]
The advantage of this ratio is that it needs no estimate of the
local mean level spacing: the energy scale cancels between numerator
and denominator, giving a simple, dimensionless
diagnostic~\cite{Oganesyan2007,Atas2013}.
Plotting its mean against the disorder strength for several
system sizes $L$ provides a way to look for the transition, though
finite-size effects can be substantial~\cite{Luitz2015}, and recent investigations have cast doubts on early conclusions~\cite{Sels2022,Morningstar2022}.








\paragraph{Entanglement entropy.}
A second diagnostic is the entanglement entropy of eigenstates~\cite{Bauer2013}.
Divide the chain into two equal parts, $A$ and $A^c$:

\begin{center}
    \begin{tikzpicture}[x=1cm,y=1cm]
        \draw (0,0) rectangle (5,0.45);
        \draw (2.5,0) -- (2.5,0.45);
        \node at (1.25,0.225) {$A$};
        \node at (3.75,0.225) {$A^c$};
    \end{tikzpicture}
\end{center}
Pick an eigenstate $|E\rangle$ and trace out $A^c$:
\[
    \rho_A(E)=\operatorname{Tr}_{A^c}|E\rangle\langle E|
    =\sum_{\phi_A,\phi'_A}|\phi_A\rangle\langle\phi'_A|
      \sum_{\phi_{A^c}}
      \langle\phi_A,\phi_{A^c}|E\rangle
      \langle E|\phi'_A,\phi_{A^c}\rangle.
\]
The entanglement entropy is
\[
    S_A(E)=-\operatorname{Tr}\bigl(\rho_A(E)\ln\rho_A(E)\bigr).
\]
MBL eigenstates retain a product structure dressed by quasi-local
correlations. They obey an area law, which in one dimension means
\[
    S_A(E)=O(1)
\]
as $L$ goes to infinity~\cite{Bauer2013}.
Thermal eigenstates instead obey a volume law. At infinite temperature, its leading term is
\[
    S_A(E)\sim\frac{\ln2}{2}L.
\]
This agrees with the leading entanglement of a random pure
state~\cite{Page1993}. 
At other energies, the coefficient $\ln 2$ is replaced by the corresponding thermal entropy density.

\section{Mechanisms for and against MBL}\label{sec: pro and con MBL}

We now describe the main mechanism that makes MBL plausible:
disorder creates frequency mismatches that suppress energy exchange.
We then turn to its main enemy, resonances, where these mismatches
are too small compared with the couplings. Resonances can potentially
trigger thermal avalanches, making MBL fragile.

\subsection{The role of disorder and resonances}

To understand whether a perturbation preserves localization or
destroys it, we need to determine how strongly it mixes the
unperturbed eigenstates. The relevant quantity is the ratio of
the matrix element coupling two states to their energy difference.
We can see this explicitly in a two-level system, or a single spin (ss):
\[
    H_{\mathrm{ss}}=\begin{pmatrix}V_1&\gamma\\ \gamma&V_2\end{pmatrix}.
\]
Here the perturbation couples the two basis states with matrix
element $\gamma>0$, and their unperturbed energy difference is
$\Delta V=V_2-V_1$.

This problem can be solved exactly. Its eigenvalues are
\[
    E_\pm=\frac{V_1+V_2}{2}
    \pm\sqrt{\left(\frac{\Delta V}{2}\right)^2+\gamma^2}.
\]
For $\Delta V\neq0$, the eigenvectors can be written as
\[
    |E_1\rangle=
    \begin{pmatrix}\cos\theta\\-\sin\theta\end{pmatrix},
    \qquad
    |E_2\rangle=
    \begin{pmatrix}\sin\theta\\\cos\theta\end{pmatrix},
    \qquad
    \tan(2\theta)=\frac{2\gamma}{\Delta V}.
\]
Thus the mixing is controlled by the ratio $\gamma/\Delta V$.

Suppose now that $V_1,V_2$ are independent random variables spread
over $[-1,1]$, and $\gamma \ll 1$.
Typically, 
$|\gamma/\Delta V|\ll1$, so the eigenvectors are, to first order and up to normalization,
\[
    |E_1\rangle\simeq
    \begin{pmatrix}1\\-\gamma/\Delta V\end{pmatrix},
    \qquad
    |E_2\rangle\simeq
    \begin{pmatrix}\gamma/\Delta V\\1\end{pmatrix},
\]
i.e.\@ they still look like
\[
    |\uparrow\rangle=\begin{pmatrix}1\\0\end{pmatrix},
    \qquad
    |\downarrow\rangle=\begin{pmatrix}0\\1\end{pmatrix}.
\]
There is little hybridization: disorder keeps the two levels sufficiently far
apart compared with their coupling.
Occasionally, however, $|\Delta V|$ is comparable to $\gamma$
or smaller. This is a resonance, and the eigenstates become
strongly hybridized. At $V_1=V_2$, they are
\[
    \frac{|\uparrow\rangle\pm|\downarrow\rangle}{\sqrt2}.
\]
Even a weak perturbation can therefore produce strong mixing
if the energy difference is small enough.

This single-spin calculation illustrates the general rule:
what matters is not the perturbation alone, but the ratio
\begin{equation}\label{eq: matrix element divided by denominator}
    \frac{|\text{matrix element of the perturbation}|}
         {|\text{unperturbed energy difference}|}.
\end{equation}
MBL will be a far-reaching many-body application of this idea.
The challenge is to control the many possible transitions and
the resonances that appear on increasingly large scales.

\subsection{Schrieffer--Wolff transformation}\label{sec: schrieffer wolff}

We now lift this idea to the many-body setting.
Consider again the Hamiltonian $H$ in \eqref{eq: main hamiltonian},
written as
\[
    H=E^{(0)}+\gamma V,
    \qquad
    E^{(0)}=\sum_{i=1}^L h_iZ_i+J\sum_{i=1}^{L-1} Z_iZ_{i+1},
    \qquad V=\sum_{i=1}^L X_i.
\]
We recall that $h_i$ are random magnetic fields, uniformly distributed in $[-1,1]$.

As a first step towards diagonalizing $H$, we construct a unitary
$U^{(1)}$ that preserves locality and removes the off-diagonal terms
to first order in $\gamma\ll 1$. This is a Schrieffer--Wolff
transformation~\cite{Schrieffer1966,Bravyi2011}. It gives an
approximate diagonalization, not yet the full change of basis.
Take
\[
    U^{(1)}=e^{-\gamma A},\qquad A^\dagger=-A.
\]
As $H$ is real and symmetric, $A$ can be chosen real and antisymmetric, so
$U^{(1)}$ is orthogonal. Expanding in powers of $\gamma$ gives
\begin{align*}
    (U^{(1)})^\dagger HU^{(1)}
    &=e^{\gamma A}(E^{(0)}+\gamma V)e^{-\gamma A}\\
    &=e^{\gamma[A,\cdot]}(E^{(0)}+\gamma V)\\
    &=E^{(0)}+\gamma V+\gamma[A,E^{(0)}]
      +\gamma^2[A,V]+\frac{\gamma^2}{2}[A,[A,E^{(0)}]]
      +O(\gamma^3).
\end{align*}
Here and throughout, the notation $O(\gamma^p)$ is simply used for power counting in $\gamma$.
To cancel the first-order term, we solve
\[
    V=[E^{(0)},A].
\]
Since $V=\sum_iX_i$, we set $A=\sum_iA_i$ and solve
\[
    X_i=[E^{(0)},A_i].
\]
The solution is
\[
    A_i=-\frac{1}{\Delta_i E^{(0)}}X_i,
    \qquad \Delta_i E^{(0)}=X_i E^{(0)} X_i - E^{(0)}.
\]
The denominator is diagonal in the classical basis and almost surely never vanishes. 

The matrix elements of $\gamma A_i$ are ratios of a perturbation
matrix element to an energy difference, as in
\eqref{eq: matrix element divided by denominator}.
The only nonzero ones connect configurations that differ by a
spin flip at site $i$. For a bulk site, 
\begin{align}
    \langle\ldots,-s_i,\ldots|\gamma A_i|\ldots,s_i,\ldots\rangle
    &=\frac{\gamma}{E^{(0)}(\ldots,-s_i,\ldots)-E^{(0)}(\ldots,s_i,\ldots)}
    \nonumber\\
    &=-\frac{\gamma}{2s_i\bigl(h_i+J(s_{i-1}+s_{i+1})\bigr)}.
    \label{ratio resonance first order}
\end{align}
Since this expression depends only on $s_{i-1},s_i,s_{i+1}$,
the operator $A_i$ acts only on site $i$ and its two neighbors.
Since $\gamma \ll 1$, at a typical site, these ratios are small in absolute value for
all configurations of the three spins, and the rotation is
perturbative. At a resonant site, at least one ratio is of order
one or larger in absolute value. Such sites are rare at weak
coupling, but occur with a nonzero density in a long chain.

For the moment, make the unrealistic assumption that resonances
never appear and that the denominators are uniformly controlled.
Then
\[
    U^{(1)}=e^{-\gamma\sum_iA_i}
\]
preserves locality, and the transformed Hamiltonian is
\[
    H^{(1)}:=(U^{(1)})^\dagger HU^{(1)}
    =E^{(0)}+\frac{\gamma^2}{2}[A,V]+O(\gamma^3).
\]
The first-order off-diagonal term has disappeared, leaving a
correction of order $\gamma^2$.

This suggests an iterative diagonalization~\cite{Imbrie2016}.
To prepare the next step, split the transformed Hamiltonian as
\[
    H^{(1)}=E^{(1)}+\gamma^2 V^{(1)},
    \qquad
    E^{(1)}=\operatorname{diag}H^{(1)}.
\]
To leading order,
\[
    E^{(1)}
    =E^{(0)}+\frac{\gamma^2}{2}\operatorname{diag}[A,V]
      +O(\gamma^3),
\]
and
\[
    V^{(1)}
    =\frac12
      \bigl([A,V]-\operatorname{diag}[A,V]\bigr)
      +O(\gamma).
\]
It is important to separate the diagonal terms: they cannot be
removed by solving a commutator equation with a diagonal Hamiltonian,
since the corresponding energy differences vanish.
They do not need to be removed either: they shift, or renormalize,
the energies and are included in the new unperturbed Hamiltonian.

We can now repeat the procedure starting from $H^{(1)}$, treating $\gamma^2V^{(1)}$ as the perturbation. 
With controlled denominators, this gives an idealized scheme in which
\[
    \gamma_{n+1}\sim\gamma_n^2,
    \qquad \gamma_0=\gamma,
    \qquad \gamma_n\sim\gamma^{2^n},
\]
where $\gamma_n$ denotes the bare perturbative parameter at step $n$. 
This resembles a Newton or KAM scheme.
We will return to it in Section~\ref{sec: math MBL}.
Before that, we need to examine the role of resonances more carefully.

\subsection{Can resonances spoil the game? Avalanches}

Let us examine a single resonant region in an otherwise localized
chain~\cite{DeRoeck2017,Luitz2017}. We keep the same Hamiltonian
as before, but imagine a disorder realization with just one
resonant spot: a consecutive set of spins for which perturbation
theory fails. A large sample will generally contain many such
spots, but we first isolate the effect of one. For simplicity,
we place the spot at the left end of the chain.

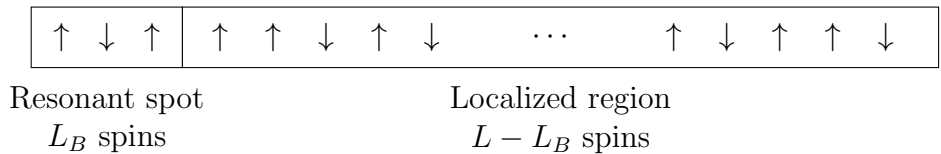
\begin{center}
\begin{tikzpicture}[x=1cm,y=1cm]
    \filldraw[fill=orange!0,draw=black!100]
        (0,0) rectangle (2,0.8);
    \foreach \x/\s in {0.4/{\uparrow},1.0/{\downarrow},1.6/{\uparrow}}
        \node at (\x,0.4) {$\s$};

    \filldraw[fill=blue!0,draw=black!100]
        (2,0) rectangle (12,0.8);
    \foreach \x/\s in {2.5/{\uparrow},3.2/{\uparrow},3.9/{\downarrow},
                       4.6/{\uparrow},5.3/{\downarrow},
                       8.5/{\uparrow},9.2/{\downarrow},
                       9.9/{\uparrow},10.6/{\uparrow},11.3/{\downarrow}}
        \node at (\x,0.4) {$\s$};
    \node at (6.9,0.4) {$\cdots$};

    \node[below,align=center] at (1,-0.1)
        {Resonant spot\\$L_B$ spins};
    \node[below,align=center] at (7,-0.1)
        {Localized region\\$L-L_B$ spins};
\end{tikzpicture}
\end{center}

Split the Hamiltonian \eqref{eq: main hamiltonian} as
\[
    H=H_B+H_{\mathrm{loc}}+H_{B-L}.
\]
Here $H_B$ acts on the resonant spot, $H_{\mathrm{loc}}$ on the
localized region, and $H_{B-L}$ couples the two.

We assume that the spot is internally thermalizing. This is a
substantial assumption: a resonance alone does not imply
thermalization. It is nevertheless plausible for a sufficiently
large region where the disorder is unusually weak, so that the
region locally resembles the thermal side of the transition.
The letter $B$ stands for bath, but this is a finite, imperfect
bath within the system. The whole chain remains closed.
Importantly, its size $L_B$ is fixed, while the localized region
grows with $L$.

The isolated Hamiltonian $H_{\mathrm{loc}}$ is assumed to be fully
MBL and hence described by LIOMs. We diagonalize it by a quasi-local
change of basis. We can also diagonalize $H_B$ separately.
The remaining question is whether the coupling $H_{B-L}$ strongly
mixes the product eigenstates of these two regions.

So far, we have only split and changed the basis of our original
Hamiltonian.
To proceed, we will assume a very schematic form for $H$, expressed in the basis where $H_B$ and $H_{\mathrm{loc}}$ have been diagonalized:
\[
    H=H_B+\sum_{i = 1}^{L-L_B}\widetilde h_i\widetilde Z_i
      +g\sum_{i=1}^{L-L_B}e^{-i/\xi}X_b\widetilde X_i.
\]
On should not attempt to derive this expression from \eqref{eq: main hamiltonian}, as it really is a toy model. 
The index $i$ labels localized spins by their distance from the
bath. The operators $\widetilde Z_i$ are the LIOMs of the isolated
localized region, and $\widetilde X_i$ flips their eigenvalues.
The fields $\widetilde h_i$ are effective fields, not necessarily
the original $h_i$.

We model $H_B$ by a $2^{L_B}\times2^{L_B}$ GOE matrix and take
$X_b$ to be a bounded operator acting on the bath. When estimating
matrix elements below, we use the eigenbasis of $H_B$.
The toy model is a bold simplification of the original model, but it retains the feature that matters here: the coupling to a localized mode decays exponentially with its distance from the original bath, with localization length $\xi$.

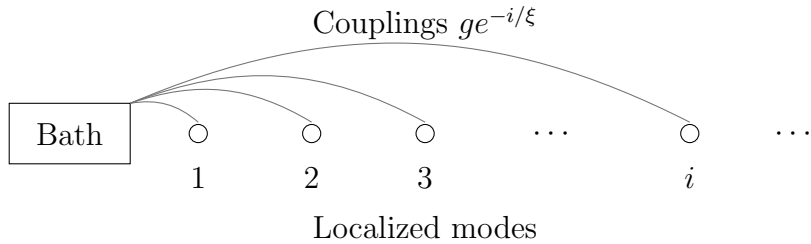
\begin{center}
\begin{tikzpicture}[x=1cm,y=1cm]
    \filldraw[fill=orange!0,draw=black!100]
        (0,0) rectangle (1.6,0.8);
    \node at (0.8,0.4) {Bath};

    \foreach \x/\i in {2.5/1,4/2,5.5/3,9/{$i$}}
    {
        \filldraw[fill=blue!0,draw=black!100]
            (\x,0.4) circle (0.12);
        \node[below] at (\x,0.1) {\i};
        \draw[black!55,bend left=25]
            (1.6,0.8) to (\x,0.55);
    }
    \node at (7.2,0.4) {$\cdots$};
    \node at (10.4,0.4) {$\cdots$};
    \node[above] at (5.5,1.5)
        {Couplings $g e^{-i/\xi}$};
    \node[below] at (5.5,-0.55) {Localized modes};
\end{tikzpicture}
\end{center}

The avalanche scenario goes as follows. The bath strongly
hybridizes the first localized spin with its own states. This
spin then becomes part of an enlarged effective bath. The bath
Hilbert space doubles, its level spacing decreases, and it can
more easily hybridize another spin. If this continues, the whole
chain thermalizes.
There is, however, a competing effect: the coupling to more
distant spins decreases as $e^{-i/\xi}$. Two exponentials therefore
compete, the decreasing bath level spacing and the decreasing
coupling strength. We need to determine which wins.

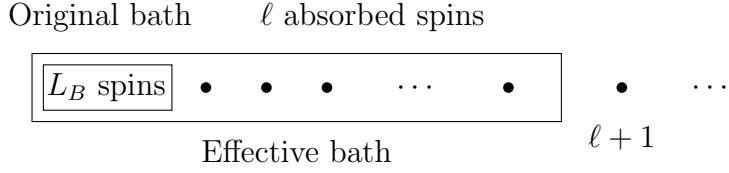
\begin{center}
\begin{tikzpicture}[x=1cm,y=1cm]
    \filldraw[fill=orange!0,draw=black!100]
        (0,0) rectangle (7,0.9);

    \filldraw[fill=orange!0,draw=black!100]
        (0.15,0.15) rectangle (1.85,0.75);
    \node at (0.98,0.45) {$L_B$ spins};
    \node[above] at (0.9,1.05) {Original bath};

    \foreach \x in {2.3,3.1,3.9,6.3}
        \fill (\x,0.45) circle (2pt);
    \node at (5.1,0.45) {$\cdots$};
    \node[above] at (4.5,1.05) {$\ell$ absorbed spins};
    \node[below] at (3.5,-0.1) {Effective bath};

    \fill (7.8,0.45) circle (2pt);
    \node[below] at (7.8,0.1) {$\ell+1$};
    \node at (9,0.45) {$\cdots$};
\end{tikzpicture}
\end{center}

Suppose the bath has already absorbed $\ell$ spins. Can it absorb
spin $\ell+1$? 
Once again, as in \eqref{eq: matrix element divided by denominator}, we compare a coupling
matrix element with the energy difference between the two states it connects.
Writing $E_b(\ell)$ for an eigenvalue of the effective bath, this ratio is
\[
    \mathcal G_\ell=
    \frac{\left|\langle E_b'(\ell),\downarrow|
        g e^{-(\ell+1)/\xi}X_b\widetilde X_{\ell+1}
        |E_b(\ell),\uparrow\rangle\right|}
    {\left|E_b'(\ell)-E_b(\ell)-2\widetilde h_{\ell+1}\right|}, 
\]
where the arrows now denote eigenstates of $\widetilde Z_{\ell+1}$.
Note that the absorption process described here is a step-by-step
construction of the eigenstates of the coupled system, not a time evolution.

We can typically find a bath state for which the denominator
is of the order of the bath level spacing.
The effective bath has dimension $2^{L_B+\ell}$ and its level spacing scales thus as $2^{-(L_B+\ell)}$.
The off-diagonal ETH estimate for the bounded operator $X_b$
gives matrix elements of order $2^{-(L_B+\ell)/2}$.
Therefore
\[
    \mathcal G_\ell\sim
    g e^{-(\ell+1)/\xi}
    \frac{2^{-(L_B+\ell)/2}}{2^{-(L_B+\ell)}}
    =g e^{-1/\xi}2^{L_B/2}
      e^{\ell(\ln2/2-1/\xi)}.
\]
If $L_B$ is sufficiently large, $\mathcal G_0\gg1$, so the bath
can hybridize nearby spins. 
The question is whether this remains possible as $\ell$ increases.
This simple estimate leads to several conclusions about the
stability of MBL and how to interpret numerical results.

\paragraph{A critical localization length.}
For the one-sided geometry considered above, the two exponentials
balance at
\[
    \xi_c=\frac{2}{\ln2}.
\]
If $\xi<\xi_c$, then $\mathcal G_\ell$ decreases and eventually
becomes small: the avalanche halts. A thermal spot then affects
only a finite neighborhood, leaving room for a stable MBL phase.
If $\xi>\xi_c$, the ratio grows, and a sufficiently large initial
bath can trigger an avalanche through the whole chain.

\paragraph{The role of geometry.}
The threshold depends on how the bath grows. If localized material
lies on both sides, absorbing $\ell$ spins on each side increases
the bath size by $2\ell$, while the coupling to the next spins
still decays as $e^{-\ell/\xi}$. The same counting then gives
\[
    \xi_c=\frac{1}{\ln2}.
\]
In dimensions $d>1$, the absorbed volume grows as a power $r^d$
of the distance $r$, whereas the coupling decays exponentially
in $r$. The growth of the bath eventually wins for any
$\xi>0$, provided a sufficiently large thermal seed is present.
The avalanche argument therefore predicts $\xi_c=0$ (no MBL) when $d>1$~\cite{DeRoeck2017}.

\paragraph{Interpreting numerical results.}
These ideas change how we interpret finite-size numerics.
Sels and Morningstar et al.\ studied chains coupled at one end
to a fixed Markovian bath, using a Lindblad
description~\cite{Sels2022,Morningstar2022}. Unlike the internal
finite bath above, this is an external reservoir, introduced
as a probe of avalanche stability.
Their results suggest that a true MBL phase, if present,
requires substantially stronger disorder than many earlier
finite-size estimates indicated. A system can therefore look
localized on accessible sizes and times while ultimately
thermalizing. 

\paragraph{The MBL--ETH transition.}
The avalanche picture constrains the transition itself.
The localization length controlling the decay of couplings
cannot diverge on approaching the transition from the localized
side: sufficiently large thermal spots would already trigger
avalanches at a finite value of $\xi$. 
This differs from the usual picture of an Anderson transition~\cite{Thiery2018}.

\section{Mathematical approach to MBL}\label{sec: math MBL}

In this last section, we follow the approach of De Roeck et al.~\cite{DeRoeck2024},
which builds on ideas developed by Imbrie~\cite{Imbrie2016}.
See also Scardicchio and Sondhi~\cite{ScardicchioSondhi2026}
for a numerical implementation of this approach.

\subsection{MBL away from resonances}

We set ourselves a restricted task: prove MBL in a
resonance-free environment, where resonant transitions never occur
at any step of the construction. Such an environment is atypical
in a long chain.

At the first perturbative step, \eqref{ratio resonance first order}
shows that a resonance occurs when
\[
    \left|\frac{\gamma}{\Delta_i E^{(0)}}\right|
    =\left|\frac{\gamma}
      {2s_i\bigl(h_i+J(s_{i-1}+s_{i+1})\bigr)}\right|
    \gtrsim1.
\]
For our regular disorder distribution, $h_i$ uniform in $[-1,1]$, these events are rare at weak
coupling but unavoidable as the length of the chain increases. 
A first-order estimate is
\[
    \mathbb P(\text{resonance-free stretch of length }\ell)
    \sim\left(1-c \gamma\right)^\ell,
\]
for some $c>0$. 
When higher-order resonances are included, the precise rate changes, but an exponential lower bound in $\ell$ can still be established~\cite{DeRoeck2024}.

Proving MBL on this event therefore establishes localization
with a probability that is exponentially small in the system
size. Why pursue such a result? There are two reasons.
First, it is a useful mathematical intermediate step: even without
resonances, the construction reveals important difficulties.
Second, rare localized stretches can have consequences for
transport that hold with overwhelming probability in a large
system.

The mechanism is a Griffiths effect: rare insulating stretches
act as bottlenecks, limiting transport even when the whole chain
is not localized~\cite{Agarwal2015,Gopalakrishnan2015}.
To probe transport, place a chain of length $L$ between hot and
cold reservoirs. 
These reservoirs drive a current into the chain.
Write $\mathcal J$ for the heat current and let $\Delta T$ be the
temperature difference.
Let us distinguish three possibilities for transport: 
\begin{enumerate}
    \item Diffusion: 
    \[
    \mathcal J\sim\frac{\Delta T}{L}.
    \]
    \item Subdiffusion: 
    \[
    \mathcal J\sim\frac{\Delta T}{L^a},\qquad a>1.
    \]
    \item MBL: 
    \[
    \mathcal J\sim\Delta T\,e^{-L/\xi}.
    \]
\end{enumerate}

An exponentially small probability of finding a resonance-free
stretch of length $\ell$ implies that the longest such stretch
in a chain of length $L$ has a typical length of order
\[
    \ell=K\log L.
\]
If this stretch is localized, it limits the current through the
whole chain. The bottleneck estimate is
\[
    |\mathcal J|\lesssim|\Delta T|\,e^{-\ell/\xi}
      =\frac{|\Delta T|}{L^{K/\xi}}.
\]
Making $\gamma$ smaller makes resonance-free stretches more
likely and localization stronger: $K$ increases and $\xi$
decreases. The exponent $K/\xi$ can therefore exceed one,
giving a subdiffusive upper bound on the current.

A rigorous version of this argument holds with overwhelming
probability as $L$ grows~\cite{DeRoeck2024}. Thus, although an
entirely resonance-free chain is rare, the transport consequences
of resonance-free stretches are typical.
We now go through the main challenges to be overcome in order to establish the localization result.

\subsection{Renormalization vs perturbation theory}

At the end of Section~\ref{sec: schrieffer wolff}, we outlined the possibility of setting up a renormalization KAM-like scheme to derive MBL. 
But let's first ask ourselves the question of why we can't simply solve this problem by applying regular perturbation theory, which would be tempting since we now exclude all resonances at a fixed value of $\gamma$.
We provide an answer at two different levels.

\paragraph{Avoided level crossings.}
A first general reason is the presence of avoided crossings.
As the discussion at the end of
Section~\ref{sec: schrieffer wolff} shows, energy levels acquire
corrections of the form
\[
    E_i(\gamma)=E_i^{(0)}+f_i\gamma^2+\cdots.
\]
The levels therefore move as $\gamma$ varies, and some approach
one another closely.

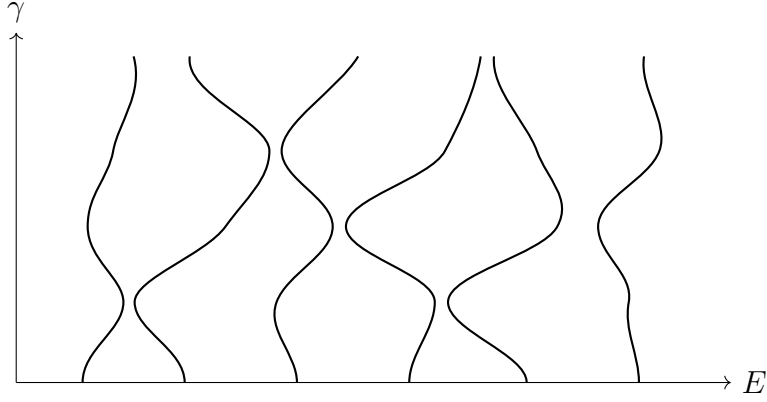
\begin{center}
\begin{tikzpicture}[x=1.35cm,y=1.25cm]
    \draw[->] (0,0) -- (7,0) node[right] {$E$};
    \draw[->] (0,0) -- (0,3.7) node[above] {$\gamma$};

    \begin{scope}[thick]
        \draw (0.65,0)
            .. controls (0.65,0.35) and (1.05,0.55) .. (1.05,0.85)
            .. controls (1.05,1.10) and (0.70,1.30) .. (0.70,1.65)
            .. controls (0.70,2.00) and (0.90,2.15) .. (0.95,2.45)
            .. controls (1.00,2.75) and (1.25,3.05) .. (1.15,3.45);

        \draw (1.65,0)
            .. controls (1.65,0.35) and (1.16,0.55) .. (1.16,0.85)
            .. controls (1.16,1.10) and (1.85,1.35) .. (2.05,1.65)
            .. controls (2.25,1.95) and (2.48,2.15) .. (2.48,2.45)
            .. controls (2.48,2.75) and (1.65,3.05) .. (1.70,3.45);

        \draw (2.75,0)
            .. controls (2.75,0.30) and (2.45,0.55) .. (2.55,0.85)
            .. controls (2.65,1.15) and (3.10,1.35) .. (3.10,1.65)
            .. controls (3.10,1.95) and (2.60,2.15) .. (2.60,2.45)
            .. controls (2.60,2.75) and (3.15,3.10) .. (3.35,3.45);

        \draw (3.85,0)
            .. controls (3.85,0.30) and (4.10,0.55) .. (4.10,0.85)
            .. controls (4.10,1.15) and (3.23,1.35) .. (3.23,1.65)
            .. controls (3.23,1.95) and (4.05,2.15) .. (4.20,2.45)
            .. controls (4.35,2.75) and (4.50,3.10) .. (4.55,3.45);

        \draw (5.00,0)
            .. controls (5.00,0.30) and (4.23,0.55) .. (4.23,0.85)
            .. controls (4.23,1.15) and (5.15,1.35) .. (5.30,1.65)
            .. controls (5.45,1.95) and (5.20,2.15) .. (5.10,2.45)
            .. controls (5.00,2.75) and (4.65,3.10) .. (4.68,3.45);

        \draw (6.10,0)
            .. controls (6.10,0.30) and (5.95,0.55) .. (6.00,0.85)
            .. controls (6.05,1.15) and (5.70,1.35) .. (5.70,1.65)
            .. controls (5.70,1.95) and (6.20,2.15) .. (6.30,2.45)
            .. controls (6.40,2.75) and (6.10,3.10) .. (6.15,3.45);
    \end{scope}
\end{tikzpicture}
\end{center}

In the bulk of the spectrum, the mean level spacing is of order $2^{-L}$ whereas the leading energy shifts are of order $\gamma^2$. 
These shifts can therefore bring neighboring levels close together even at very small
$\gamma$, but a nonzero coupling between the levels generically turns a crossing into an
avoided crossing.

The important point is that an avoided crossing on the real axis is typically accompanied by eigenvalue branch points nearby in the complex $\gamma$ plane. 
Therefore, the eigenvalues can remain analytic along the real axis while their Taylor series
about $\gamma=0$ have a radius of convergence that decays exponentially with system size.
This explains why we cannot proceed with a convergent perturbative expansion: 
reaching larger real values of $\gamma$ requires analytic continuation.

Let us make this argument more explicit.
Consider two nearby levels, described by the effective matrix
\[
    \begin{pmatrix}
        V-t(\gamma)&b(\gamma)\\
        b(\gamma)&-V+t(\gamma)
    \end{pmatrix},
    \qquad
    E_\pm(\gamma)=
    \pm\sqrt{(V-t(\gamma))^2+b(\gamma)^2}.
\]
We have omitted a common energy shift, which does not affect
the crossing. Take $V>0$, with initial level separation
\[
    2V\sim2^{-L}.
\]
To describe levels moving towards one another, take
\[
    t(\gamma)=a\gamma^2+\cdots,\qquad a>0,
\]
with $a$ of order one. Without the coupling $b$, the levels
would cross at
\[
    \gamma_*\simeq\sqrt{V/a}.
\]

In a localized system, nearby many-body levels typically
correspond to configurations differing over a large part of
the chain. Their coupling involves high-order processes and
can be exponentially small in $L$:
\[
    |b(\gamma_*)|=O(e^{-cL/\xi}),
    \qquad 0<c<1.
\]
We consider sufficiently small $\xi$ that $1/\xi\gg\ln2$,
so this coupling is much smaller than the initial spacing $V$.

The eigenvalues have branch points where
\[
    V-t(\gamma)=\pm i b(\gamma).
\]
Linearizing near $\gamma_*$ gives, to leading order,
\[
    \gamma_{\mathrm{bp}}
    \simeq\gamma_*
      \pm\frac{i b(\gamma_*)}{2a\gamma_*}.
\]
The branch points therefore lie very close to the real avoided
crossing, at a distance from the origin of order
\[
    \gamma_{\mathrm{bp}}
    \sim\gamma_*
    \sim\sqrt V
    \sim2^{-L/2}.
\]
Such a crossing limits the radius of convergence of the Taylor
series about $\gamma=0$ to an exponentially small scale.
This gives a heuristic obstruction to a purely perturbative
expansion converging in a neighborhood whose size is independent
of $L$.

Finally, how does this fit with the absence of level repulsion
discussed in Section~\ref{sec: numerical probe MBL}?
In a localized system, the avoided crossings occur on such small energy scales that,
at fixed $\gamma$ and sufficiently strong disorder, a typical
sample may show no visible trace of them~\cite{Morningstar2022}.
They become apparent only when $\gamma$ is finely tuned to bring
two levels close enough together. This is compatible with
Poisson level statistics at fixed $\gamma$.

\paragraph{Factorial growth of the number of terms.}
A more direct way to understand the breakdown of the perturbative
expansion is to examine the number of terms it generates.
As we will see, this number grows factorially with the order,
so estimating each term in absolute value would not yield a convergent expansion~\cite{Sirker2026}.

To see this, we start as in Section~\ref{sec: schrieffer wolff}, but we never renormalize the energy after the first step.
Concretely, suppose that after $n\ge 1$ steps we have
\[
    H^{(n)}=E^{(0)}+\gamma^2D^{(n)}+\gamma_nV^{(n)},
\]
where $D^{(n)}$ is diagonal, $V^{(n)}$ is off-diagonal, and
$\gamma_n$ is the running coupling constant.
As explained in Section~\ref{sec: schrieffer wolff}, diagonal terms cannot be rotated away and this is why the term $\gamma^2 D^{(n)}$ appears here. 
If we keep using the original energies, without renormalization,
we choose a generator $A^{(n+1)}$ solving
\begin{equation}\label{eq: commutator equation perturbative}
    [E^{(0)},A^{(n+1)}]=V^{(n)},
\end{equation}
and set
\[
    H^{(n+1)}
    =e^{\gamma_n[A^{(n+1)},\cdot]}H^{(n)}.
\]
Expanding gives
\begin{align*}
    H^{(n+1)}
    &=E^{(0)}+\gamma^2D^{(n)}
      +\gamma_n\bigl(V^{(n)}+[A^{(n+1)},E^{(0)}]\bigr)\\
    &\quad+\gamma^2\gamma_n[A^{(n+1)},D^{(n)}]
      +O(\gamma_n^2)\\
    &=E^{(0)}+\gamma^2D^{(n)}
      +\gamma^2\gamma_n[A^{(n+1)},D^{(n)}]
      +O(\gamma_n^2).
\end{align*}
(note that for $n=1$, the remainder is of the same order as the displayed correction, but this is not the case for $n>1$).
We see that the term of order $\gamma_n$ cancels, but commuting with the energy correction produces a new off-diagonal term of order $\gamma^2 \gamma_n$.
All newly generated diagonal terms will be absorbed into $D^{(n+1)}$.

Thus, starting with $\gamma_1\sim\gamma^2$, the powers of
$\gamma$ improve as
\[
    \gamma_{n+1}\sim\gamma^2\gamma_n,
    \qquad \gamma_n\sim\gamma^{2n}.
\]
This is slower than the quadratic improvement obtained by
including the energy corrections in the unperturbed Hamiltonian in the KAM-like scheme.

Meanwhile, the number of terms grows.
At leading order $\gamma^{2n}$, each local term in $V^{(n)}$ involves $2n$ transverse-field operators, whose sites lie within an interval of at most $2n$ sites.
The energy denominators, introduced by solving \eqref{eq: commutator equation perturbative}, can depend additionally on one neighboring spin at each end.
Thus the corresponding terms in $A^{(n+1)}$ have support on at most $2(n+1)$ sites.
For the following schematic count, we use this upper bound on their support length.
Now write
\[
    [A^{(n+1)},D^{(n)}]
    =\sum_{i,j}[A_i^{(n+1)},D_j^{(n)}].
\]
For a fixed $A_i^{(n+1)}$, only overlapping terms
$D_j^{(n)}$ can contribute. Considering just the short-range
terms in the diagonal correction, there are about $2(n+1)$
possible positions:

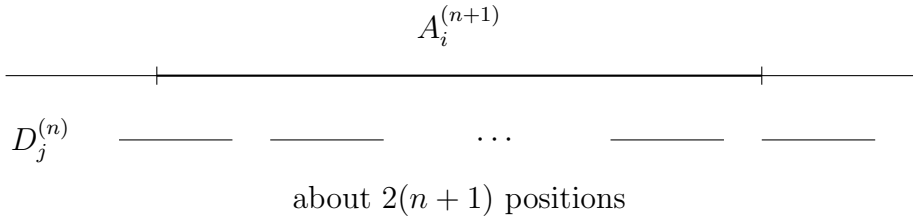
\begin{center}
\begin{tikzpicture}[x=1cm,y=1cm]
    \draw (0,0) -- (12,0);
    \draw[thick] (2,0) -- (10,0);
    \draw (2,-0.12) -- (2,0.12);
    \draw (10,-0.12) -- (10,0.12);
    \node[above] at (6,0.25) {$A_i^{(n+1)}$};

    \draw (1.5,-0.85) -- (3,-0.85);
    \draw (3.5,-0.85) -- (5,-0.85);
    \node at (6.5,-0.85) {$\cdots$};
    \draw (8,-0.85) -- (9.5,-0.85);
    \draw (10,-0.85) -- (11.5,-0.85);
    \node[left] at (1,-0.85) {$D_j^{(n)}$};

    \node at (6,-1.65) {about $2(n+1)$ positions};
\end{tikzpicture}
\end{center}
The number $R_n$ of local terms is therefore multiplied at each step
by a factor proportional to $n$: 
\[
    R_{n+1} \sim 2(n+1) R_n, \qquad R_n \sim 2^n n!
\]
The exponential decay of $\gamma_n \sim \gamma^{2n}$ will not be able to make up for this exponential growth. 


As a side note, let us mention that the same factorial growth appears in perturbative approaches
to prethermalization~\cite{Mori2016,Abanin2017Prethermal,Abanin2017Rigorous}.
There, truncating the expansion at a suitable order gives an
approximately conserved quantity and a long-lived prethermal
regime. In a generic thermalizing system, the breakdown of the expansion is
expected to reflect the true behavior of the system: the approximately conserved quantity is eventually destroyed as the system thermalizes. 
MBL goes beyond prethermalization: by renormalizing the energies
at each step, we seek a convergent construction of exact
quasi-local conserved quantities.

\subsection{Renormalization scheme}

We now return to the renormalization procedure outlined in Section~\ref{sec: schrieffer wolff}. 
By updating the unperturbed energies at each step, we recover a quadratic decrease of the perturbation (note though that the scheme in \cite{Imbrie2016} and \cite{DeRoeck2024} is not exactly quadratic due to the presence of small denominators).
As we show now, this scheme removes the combinatorial obstruction discussed above. This provide thus a first step towards proving convergence.

The starting point is unchanged:
\[
    H^{(0)}=E^{(0)}+\gamma V^{(0)},
    \qquad [E^{(0)},A^{(1)}]=V^{(0)}.
\]
After the first rotation, we write
\[
    H^{(1)}=E^{(1)}+\gamma^2V^{(1)},
\]
where $E^{(1)}=\operatorname{diag}H^{(1)}$ includes the full
diagonal correction and $V^{(1)}$ is off-diagonal. To leading
order,
\[
    E^{(1)}=E^{(0)}+\gamma^2D^{(1)}+O(\gamma^3),
    \qquad
    D^{(1)}=\frac12\operatorname{diag}[A^{(1)},V^{(0)}].
\]
The important point is that we now use $E^{(1)}$, rather than
$E^{(0)}$, to construct the next rotation.

Suppose that, after $n\ge 1$ steps,
\[
    H^{(n)}=E^{(n)}+\gamma_nV^{(n)}
\]
where $\gamma_n$ is the running coupling constant (and $\gamma_1 = \gamma^2$).
We choose an off-diagonal generator satisfying
\[
    [E^{(n)},A^{(n+1)}]=V^{(n)}
\]
and perform the rotation
\[
    H^{(n+1)}
    =e^{\gamma_n[A^{(n+1)},\cdot]}H^{(n)}.
\]
For classical configurations $\bm s\neq \bm s'$, the generator has
matrix elements
\begin{equation}\label{eq: matrix element generator}
    \langle \bm s'|A^{(n+1)}|\bm s\rangle
    =
    \frac{\langle \bm s'|V^{(n)}| \bm s\rangle}
         {E^{(n)}(\bm s')-E^{(n)}(\bm s)}.
\end{equation}
For now, we assume that the denominators can be controlled.

Let us check what remains after the rotation. Using
$[A^{(n+1)},E^{(n)}]=-V^{(n)}$ in the full expansion gives
\[
    H^{(n+1)}
    =
    E^{(n)}
    + 
    \frac{\gamma_n^2}{2}[A^{(n+1)},V^{(n)}] + O(\gamma_n^3).
\]
Note that the dependence on $\gamma$ is only partially explicit in this formula, since $A^{(n+1)}$ and $V^{(n)}$ both depend on $\gamma$.
We absorb the full diagonal part into $E^{(n+1)}$ and write
the remaining off-diagonal part as
$\gamma_{n+1}V^{(n+1)}$: 
\[
    H^{(n+1)} = E^{(n+1)} + \gamma_{n+1} V^{(n+1)}.
\]
Thus, counting powers of $\gamma$, we find 
\[
    \gamma_{n+1}=\gamma_n^2,
    \qquad \gamma_n=\gamma^{2^n}.
\]

Let us now analyze the growth of the number of terms.
Write
\[
    V^{(n)}=\sum_i V_i^{(n)},
    \qquad A^{(n+1)}=\sum_i A_i^{(n+1)}.
\]
For this schematic argument, we assume that solving the
commutator equation enlarges the support by at most a fixed
number of sites. For the original nearest-neighbor energies,
these are the neighboring spins entering the energy denominator.
For the renormalized energies, locality requires further
analysis, see~\cite{DeRoeck2024}.

The leading new terms in $V^{(n+1)}$ come from
\[
    [A^{(n+1)},V^{(n)}]
    =\sum_{i,j}[A_i^{(n+1)},V_j^{(n)}].
\]
Only overlapping supports contribute, and the resulting
term is supported on their union:

\begin{center}
\begin{tikzpicture}[x=1cm,y=1cm]
    \draw (0,0.6) -- (6,0.6);
    \draw (0,0.48) -- (0,0.72);
    \draw (6,0.48) -- (6,0.72);
    \node[above] at (3,0.8) {$A_i^{(n+1)}$};

    \draw (4,-0.6) -- (10,-0.6);
    \draw (4,-0.72) -- (4,-0.48);
    \draw (10,-0.72) -- (10,-0.48);
    \node[below] at (7,-0.8) {$V_j^{(n)}$};
\end{tikzpicture}
\end{center}

This gives estimates for both the range and the number of
terms. 
If $\ell_n$ bounds the support length of the perturbation terms at step $n$,
then
\[
    \ell_{n+1}\leq 2\ell_n+c_0
\]
for a fixed constant $c_0$. 
Hence $\ell_n$ grows schematically as $2^n$, but any exponential growth is fine for the present argument. 
If $R_n$ counts the local perturbation terms per site,
there are $R_n$ choices from each operator and at most
$C_0\ell_n$ relative positions with overlapping supports.
Thus, within this schematic count,
\[
    R_{n+1}\leq C_0\ell_n R_n^2.
\]

Iterating gives
\[
    R_n\leq
    R_0^{2^n}
    \prod_{k=0}^{n-1}(C_0\ell_k)^{2^{n-1-k}}.
\]
To estimate this product, take logarithms and divide by $2^n$:
\[
    \frac{\log R_n}{2^n}
    \leq \log R_0
      +\sum_{k=0}^{n-1}
        \frac{\log(C_0\ell_k)}{2^{k+1}}.
\]
Since $\ell_k\leq ab^k$ for fixed constants $a,b$,
\[
    \log(C_0\ell_k)\leq\log(C_0a)+k\log b.
\]
The sum therefore remains bounded, since
$\sum_{k\geq0}\frac{1+k}{2^{k+1}}<\infty$.
Consequently, for some constant $C$,
\[
    R_n\leq C^{2^n}.
\]

Combining this with the perturbative factor gives
\[
    \gamma_nR_n\leq(C\gamma)^{2^n}\longrightarrow0
\]
for sufficiently small $\gamma$.
The rapid decrease of the perturbation therefore controls
the combinatorial growth. 
This is a familiar mechanism behind the convergence of KAM schemes.

\subsection{Energy denominators}

We have seen how the renormalization scheme controls the
proliferation of terms. We must now account for the energy
denominators. Absence of resonances cannot mean that every
denominator is of order one: as the support grows, there are
exponentially many possible transitions, and some energy
differences will be small. We will thus allow denominators
to decrease with the scale.

Recall that the matrix elements of $A^{(n+1)}$ are obtained
from those of $V^{(n)}$ by dividing by the corresponding
energy differences, as in \eqref{eq: matrix element generator}.
For a transition from a classical state $|\bm s\rangle$
to $|\bm s'\rangle$, write
\[
    \Delta E^{(n)}(\bm s,\bm s')
    =E^{(n)}(\bm s')-E^{(n)}(\bm s).
\]
Our basic non-resonance condition is
\[
    |\Delta E^{(n)}(\bm s,\bm s')|\geq\varepsilon_n.
\]
The threshold  $\varepsilon_n$ must be small enough that violations are rare,
but large enough to control division by the denominator.

As in the previous subsection, $\ell_n$ represents the
support scale of terms at step $n$. For the following
schematic calculation, we use
\[
    \ell_n=2^n,
    \qquad
    \varepsilon_n=\varepsilon^{\ell_n},
\]
for $\varepsilon$ such that $0 < \gamma \ll \varepsilon \ll 1$. 
Why this choice for $\varepsilon_n$?
On the one hand, Poisson statistics suggest that the smallest
level spacing in a region of size $\ell_n$ is typically of
order $2^{-2\ell_n}$. Choosing
$\varepsilon$ sufficiently small makes the threshold
$\varepsilon^{\ell_n}$ much smaller than this scale, so
violations of the non-resonance condition are expected to be rare.  
On the other hand, $\varepsilon_n$ is much larger than the
bare perturbative factor
$\gamma_n\sim\gamma^{2^n}=\gamma^{\ell_n}$, so division by one
such denominator still leaves a small quantity.

However, denominators accumulate at successive steps, so
the bare factor $\gamma_n$ alone does not determine the
perturbation strength.
Let us now see what happens if we bound every denominator
by its threshold. The leading term in $V^{(n+1)}$ comes from the commutator
$[A^{(n+1)},V^{(n)}]$ which combines two terms from scale $n$.
The term coming from $A^{(n+1)}$ carries one additional
denominator at that scale. Schematically, a
contribution $T^{(n+1)}$ to $\gamma_{n+1}V^{(n+1)}$ therefore has the form
\[
    T^{(n+1)}
    \sim \frac{T_1^{(n)}T_2^{(n)}}{\Delta E^{(n)}},
\]
where the $T_i^{(n)}$ include their powers of $\gamma$.

Repeating this decomposition gives a binary tree.
A term at scale $n$ contains $2^n$ elementary perturbations,
each carrying a factor $\gamma$. It also contains one
denominator at scale $n-1$, two at scale $n-2$, and so on,
down to $2^{n-1}$ denominators at scale zero.
The bound obtained by replacing each denominator by its
threshold is therefore
\[
    |T^{(n)}|
    \lesssim
    \frac{\gamma^{2^n}}
    {\displaystyle\prod_{k=0}^{n-1}
        \varepsilon_k^{2^{n-1-k}}}.
\]
Since $\varepsilon_k=\varepsilon^{2^k}$, every scale
contributes the same power of $\varepsilon$:
\[
    \prod_{k=0}^{n-1}\varepsilon_k^{2^{n-1-k}}
    =
    \varepsilon^{\sum_{k=0}^{n-1}2^k2^{n-1-k}}
    =
    \varepsilon^{n2^{n-1}}.
\]
Hence
\[
    |T^{(n)}|
    \lesssim
    \left(\frac{\gamma}{\varepsilon^{n/2}}\right)^{2^n}.
\]
For any fixed $\gamma>0$, this bound eventually grows,
so convergence of the scheme is not established.

This difficulty, and the idea for overcoming it, already
appear in Imbrie's work~\cite{Imbrie2016}.
The estimate is too pessimistic: it treats every denominator
as if it were close to its smallest allowed value.
Instead of applying this worst-case bound separately at
each step, we must, for suitable diagrams, expand back down
to scale zero and estimate the full product of denominators.

Fractional moments prove useful here~\cite{Aizenman1993,Imbrie2016}.
To illustrate the idea, consider a fixed diagram amplitude
with $m\leq\ell_n$ denominators. For $0<\alpha<1$,
Markov's inequality gives
\begin{align*}
    \mathbb P\left(
      \frac{\gamma^{\ell_n}}
           {\prod_{j=1}^{m}|\Delta E_j|}
      \geq
      \frac{\gamma^{\ell_n}}{\varepsilon^{\ell_n}}
    \right)
    &=
    \mathbb P\left(
      \prod_{j=1}^{m}|\Delta E_j|^{-\alpha}
      \geq\varepsilon^{-\alpha\ell_n}
    \right)\\
    &\leq
    \varepsilon^{\alpha\ell_n}
    \mathbb E\left(
      \prod_{j=1}^{m}|\Delta E_j|^{-\alpha}
    \right).
\end{align*}
Suppose for a moment that the denominators are independent
and satisfy
\[
    \mathbb E\bigl(|\Delta E_j|^{-\alpha}\bigr)
    \leq C_\alpha,
    \qquad C_\alpha\geq1.
\]
Such a negative moment is finite for a random variable
with a bounded density near zero, because
$\int_0^1 x^{-\alpha}\,dx=\frac{1}{1-\alpha}<\infty$.
Independence would then give
\[
    \mathbb E\left(
      \prod_{j=1}^{m}|\Delta E_j|^{-\alpha}
    \right)
    \leq C_\alpha^m
    \leq C_\alpha^{\ell_n},
\]
and consequently
\[
    \mathbb P\left(
      \frac{\gamma^{\ell_n}}
           {\prod_{j=1}^{m}|\Delta E_j|}
      \geq
      \left(\frac{\gamma}{\varepsilon}\right)^{\ell_n}
    \right)
    \leq
    \bigl(C_\alpha\varepsilon^\alpha\bigr)^{\ell_n}.
\]
For sufficiently small $\varepsilon$, the probability
decays exponentially with $\ell_n$.
Unlike the previous worst-case estimate, there is no
extra loss proportional to the number of scales.
The actual denominators are correlated, so their negative
moments cannot simply be multiplied. A rigorous implementation
combines estimates on suitable products of denominators with
inductive bounds for the remaining parts of the diagrams,
as in~\cite{DeRoeck2024}.

\subsection{Outlook}

The ideas described above can be turned into rigorous
mathematics~\cite{DeRoeck2024}. They yield a proof of MBL
on a set of disorder realizations whose probability is
bounded below by an exponentially small quantity in the
system size. Through the rare-region argument discussed
earlier, they also give a subdiffusive upper bound on
transport with probability tending to one as the chain
becomes large.

Can the MBL result itself be extended to hold with high
probability? In view of the avalanche picture, we strongly
expect this to be possible in one dimension at sufficiently
strong disorder, at least under a limited level attraction
(LLA) assumption controlling the probability of anomalously
small level spacings~\cite{Imbrie2016}.
The perturbative part is now well understood, and the
avalanche picture explains how a sufficiently localized
surrounding region can contain the effect of a thermal spot.
This picture has gained rigorous support from the proof
of localization and Poisson statistics in the quantum sun
model for sufficiently rapidly decaying
couplings~\cite{DeRoeckHannani2025}.

Combining these ingredients into a proof for a chain with
many resonant regions remains a substantial mathematical
challenge. Nevertheless, the overall strategy appears
clear: control the non-resonant regions perturbatively
and show that resonant regions cannot trigger unbounded
avalanches.

\paragraph{Use of AI.}
I used ChatGPT to help convert my original handwritten notes into this document, check some calculations and statements, and find additional references.

\paragraph{Acknowledgements.}
I warmly thank Serena Cenatiempo, Sven Bachmann, Alain Joye,
and Manfred Salmhofer for organizing the Les Houches Summer
School, \emph{Quantum Theory on All Scales}, and for the
invitation to give this mini-course. 
I am also grateful to Wojciech De Roeck for our long-standing collaboration on
this topic and for countless invaluable discussions.

\bibliographystyle{unsrt}
\bibliography{mbl_references}

\end{document}